\documentclass{knawproc2}

\input epsf
\let\sec=\section
\let\ssec=\subsection

\newcount\japif
\def\japeqn{\ifnum\japif=1
\begin{equation}\global\japif=0 \else
\end{equation}\global\japif=1\fi}

\def\ref{\bibitem[]{}\hglue -1.0ex}
\def\gs{\mathrel{\lower0.6ex\hbox{$\buildrel {\textstyle >}
 \over {\scriptstyle \sim}$}}}
\def\ls{\mathrel{\lower0.6ex\hbox{$\buildrel {\textstyle <}
 \over {\scriptstyle \sim}$}}}
\newcount\equationo
\font\japrm =  cmr10 at 10truept
\font\japit =  cmti10 at 10truept

\def\kms{\;{\rm km\,s^{-1}}}
\def\kmsmpc{\;{\rm km\,s^{-1}\,Mpc^{-1}}}

\def\mpcoh{\,h^{-1}\,{\rm Mpc}}

\def\ss{\scriptscriptstyle\rm}

\def\japfig#1#2#3{
\ifnum #2 = 1 
\begin{figure}[!t]
\epsfxsize=9.0cm
\centerline{\epsfbox[28 186 488 590]{jap_knaw_fig#1.eps}}
\fi
\ifnum #2 = 2 
\begin{figure}[!t]
\epsfxsize=12.0cm
\centerline{\epsfbox[45 286 535 500]{jap_knaw_fig#1.eps}}
\fi
\ifnum #2 = 3 
\begin{figure}[!p]
\epsfxsize=6.2cm
\epsfysize=11.3cm
\centerline{\epsfbox[85 5 460 790]{jap_knaw_fig#1.eps}}
\fi
\caption{#3}
\end{figure}
}

\def\apj{ApJ}
\def\apjs{ApJS}
\def\mn{MNRAS}

\begin{document}

\begin{opening}

\title{Radio galaxies and structure formation
\vglue -3.6cm
\centerline{\japrm  To appear in proceedings of the KNAW colloquium}
\centerline{\japit The most distant radio galaxies\/, \japrm Amsterdam, October 1997}
\vglue 2cm
}

\author{J.A. Peacock}
\addresses{%
  Royal Observatory, Blackford Hill, Edinburgh EH9 3HJ, UK\\
}
\runningtitle{Radio galaxies and structure formation}
\runningauthor{Peacock}

\end{opening}


\begin{abstract}
\noindent
This review discusses three ways in which radio galaxies and other
high-redshift objects can give us information on the nature and
statistics of cosmological inhomogeneities, and how they have evolved
between high redshift and the present: (1) The present-day
spatial distribution and clustering of radio galaxies;
(2) The evolution of radio-galaxy clustering and biased 
clustering at high redshift;
(3) Measuring density perturbation spectra from the
abundances of high-redshift galaxies.
\end{abstract}


\pretolerance 10000

\sec{Present-day clustering of radio galaxies}

Radio galaxies are interesting probes of large-scale
structure in the universe, and give a view of the
galaxy distribution which differs significantly 
from that obtained in other wavebands. Radio selection is
uniform over the sky and independent of galactic
extinction, so that reliably complete catalogues can
be obtained over large areas. As a result, the apparent
uniformity of the distribution of radio sources in early
surveys such as 4C and Parkes gave the first convincing
evidence for the large-scale homogeneity of the Universe
(Webster 1977). This uniformity arises because even
relatively bright samples of radio galaxies are at redshifts
$z\simeq 1$, so that projection effects give a huge dilution
of any intrinsic spatial clustering.
The 3D clustering of radio galaxies was first detected 
by Peacock \& Nicholson (1991; PN91), using a redshift
survey of 329 galaxies with $z<0.1$ and $S({\rm 1.4~GHz})\gs 500~{\rm mJy}$.
The result was a correlation function measured in redshift
space of 
\japeqn
\xi_{rg}(s) = [s/11 \mpcoh]^{-1.8}
\japeqn
($h\equiv H_0/100 \kmsmpc$).
This corresponds to a clustering amplitude intermediate between
normal galaxies and rich clusters of galaxies, as seems appropriate
give that radio galaxies are normally found in moderately
rich groups (e.g. Allington-Smith et al. 1993).

\japfig{1}{2}
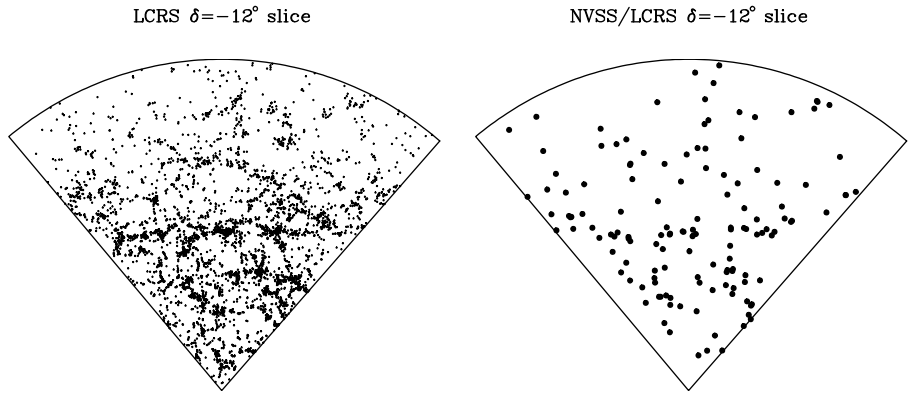
{One of the 6 slices of the Las Campanas Redshift Survey,
containing 4706 galaxies brighter than about $R=17.5$, together
with the 138 galaxies detected in the NVSS to a flux limit of
2.5 mJy at 1.4 GHz. It is clear that the radio galaxies trace out very
much the same large-scale network of filaments and
voids, albeit in a more dilute fashion.}

However, redshift surveys move on rapidly: in the region
of $10^5$ galaxy redshifts are now known, and it should be
possible to do very much better than PN91 today.
Particularly, the PN91 survey has a very low density
owing to its relatively bright flux limit; in order
to see how radio galaxies follow the pattern of
LSS in detail, deeper surveys are needed.
Rather than attempt to construct a completely new
radio-selected redshift survey, it is most efficient to
match large optically-selected redshift surveys with the
new deep all-sky radio databases such as the NVSS
and FIRST. The largest complete redshift survey for which
this exercise is possible is the Las Campanas Redshift
Survey (Shectman et al. 1996), which contains 26,418 galaxies
to $R\simeq 17.5$. The LCRS contains six slices approximately
$1.5^\circ \times 60^\circ$, of which four are at sufficiently
high declination to overlap with the NVSS
survey (Condon et al. 1997), which has a flux limit of about
2.5~mJy at 1.4~GHz. Matching these two catalogues with
a positional tolerance of 20 arcsec yields a
total of 451 radio-selected galaxies.
This significantly exceeds the PN91 sample in size,
although the relatively modest detection rate shows that
much larger parent redshift surveys such
as Sloan and 2dF will be needed to push radio-selected
redshift surveys into the $N=10^4$ -- $10^5$ regime. 
In the meantime, it is interesting to
compare the clustering in this new sample with that of PN91.

It is common to use the homogeneous distance-redshift
relation to deduce radii, and hence 3D coordinates for
galaxies in redshift surveys. From these, it is natural
to evaluate the spatial two-point correlation function $\xi(r)$
as a measure of clustering in the universe.
However, in practice the radii are affected by
large-scale coherent velocities, small-scale virialized
motions and redshift errors. A better statistic, which is independent
of all these effects, is the projected correlation function:
\japeqn
\Xi(r)=\int_{-\infty}^\infty \xi[(r^2+x^2)^{1/2}]\; dx.
\japeqn
Fig. 2 shows the result for $\Xi(r)$ measured from the
sample of 451 NVSS/LCRS galaxies.
The plot shows both the raw data, together with 
power-law models:
\japeqn
\xi(r) = (r/r_0)^{-\gamma}.
\japeqn
The canonical numbers for blue-selected radio-quiet galaxies
are $r_0=5\mpcoh$ and $\gamma=1.8$. The radio-loud result
in this sample is approximately $6.5 \mpcoh$:
the radio-loud subset of the LCRS is clustered only
slightly more strongly than the whole LCRS -- in
apparent conflict with PN91.
The larger PN91 result of $11\mpcoh$ was based on
the redshift-space $\xi(s)$, which will tend to give a
boosted value of $r_0$. In the linear analysis of
Kaiser (1987), this increases $\xi$ by a factor
\japeqn
\xi \rightarrow \xi \; [1+2\beta/3+\beta^2/5],
\japeqn
where $\beta=\Omega^{0.6}/b$ and $b$ is the bias parameter.
The relative bias between PN91 and blue-selected
galaxies gives $\beta_{\rm opt}/\beta_{\rm rad}\simeq 1.9$ 
(see Peacock \& Dodds 1994). Since $\beta_{\rm opt}$ is
currently thought to lie close to 0.5
(Hamilton 1997), this suggests
a boost factor of 1.2, and a real-space $r_0\simeq 10\mpcoh$
for the PN91 sample. This is larger than the NVSS/LCRS
result, but it is not clear that there is any
inconsistency: although the typical redshifts of the
two samples are similar, their radio flux-density
limits differ by about a factor of 200. The PN91 radio
galaxies are luminous AGN, whereas many of the
NVSS/LCRS galaxies will be starburst galaxies similar to those
found in the IRAS surveys. Starbursts obey the approximate
rule $S_{\rm 60\mu m}=90 S_{\rm 1.4~GHz}$, so that
the NVSS galaxies should be comparable in flux to
the IRAS FSS limit of 0.25~Jy
(Rowan-Robinson et al. 1994).
Since IRAS galaxies are well known
to cluster even less strongly than optical galaxies, the
modest value of $r_0$ is not a surprise.
Independent confirmation of the PN91 result for `proper'
radio galaxies will require a new radio-selected redshift survey
at brighter flux densities.

\japfig{2}{1}
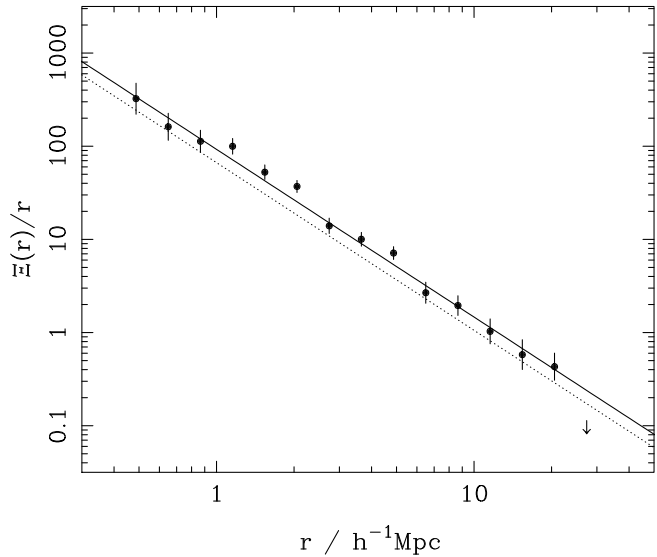
{The real-space projected correlation function, $\Xi(r)$,
for the NVSS galaxies in the four LCRS slices for which 
the two surveys overlap. The error bars indicate Poissonian
errors based on the observed pair counts, and evidently
these are an underestimate of the true error in the
intermediate regime, $r\simeq 3\mpcoh$.
The dashed line shows the normal
fit to optical galaxy clustering: $\xi=[r/r_0]^{-\gamma}$,
with $\gamma=1.8$ and $r_0=5 \mpcoh$.
The solid line shows the fit to the NVSS/LCRS sample,
indicating a slightly larger scale-length of $6.5 \mpcoh$.}

\sec{Evolution of clustering}

It would be interesting to extend these studies to higher
redshifts. This can be done without using a complete
faint redshift survey, by using the angular clustering of
a flux-limited survey. If the form of the redshift distribution
is known, the projection effects can be disentangled in order to
estimate the 3D clustering at the average redshift of the sample.
For small angles, and where the redshift shell being
studied is thicker than the scale of any clustering,
the spatial and angular correlation functions
are related by Limber's equation
(e.g. Peebles 1980):
\japeqn
w(\theta)= \int_0^\infty y^4\phi^2F^2(y)\, dy\ \int_{-\infty}^\infty
  \xi(\sqrt{x^2+y^2\theta^2})\; dx,
\japeqn
where $y$ is dimensionless comoving distance (transverse part of the FRW
metric is $[R(t) y\, d\theta]^2$), and $F(y)=[1-ky^2]^{-1/2}$;
the selection function for radius $y$ is normalized
so that $\int y^2\phi(y)F(y)\; dy=1$.

Until recently, this equation was of somewhat academic
interest for radio astronomers, since there were no
reliable detections of angular clustering. 
This has changed with the FIRST survey, which has
measured $w(\theta)$ to high precision 
for a limit of 1~mJy at 1.4~GHz
(Cress et al. 1996). 
Their result detects clustering at separations between
0.02 and 2 degrees, and is fitted by a power law:
\japeqn
w(\theta) = 0.003\, [\theta/{\rm degrees}]^{-1.1}.
\japeqn
There had been earlier claims of detections of angular
clustering, notably the 87GB survey
(Loan, Lahav \& Wall 1996), but these were of only bare
significance (although, in retrospect, the level
of clustering in 87GB is consistent with the FIRST measurement).

Limber's equation requires the redshift-dependent correlation
function, and this is commonly parameterized as follows:
\japeqn
\xi(r,z) = [r/r_0]^{-\gamma}\, (1+z)^{-(3-\gamma+\epsilon)},
\japeqn
where $\epsilon=0$ is stable clustering; $\epsilon=\gamma-3$ is
constant comoving clustering; $\epsilon=\gamma-1$ is 
$\Omega=1$ linear-theory evolution. 
Peacock (1997) showed that the expected evolution in the
quasilinear regime ($\xi \sim 1$ -- 100) is significantly
more rapid: up to $\epsilon \simeq 3$.

Discussion of the 87GB and FIRST results in terms of
Limber's equation has tended to focus on values
of $\epsilon$ in the region of 0. Cress et al. (1996) concluded that
the $w(\theta)$ results were consistent with 
the PN91 value of $r_0\simeq 10\mpcoh$ (although they were not
very specific about $\epsilon)$.
Loan et al. (1996) measured $w(1^\circ)\simeq 0.005$ for a
5-GHz limit of 50 mJy, and inferred $r_0\simeq 12\mpcoh$ for
$\epsilon=0$, falling to $r_0\simeq 9\mpcoh$ for
$\epsilon=-1$. If we take the NVSS/LCRS value
of the local clustering for radio sources,
$r_0\simeq 6.5\mpcoh$, then the observed
angular clustering in fact requires $\epsilon$ in the
region of $-1.5$ or smaller: in other words, little or
no evolution of $\xi$ with redshift.

Indeed, this conclusion can be reached in a rather
less model-dependent fashion. The reason there is
a strong degeneracy between $r_0$ and $\epsilon$ is
that $r_0$ parameterizes the $z=0$ clustering, whereas
the observations refer to a typical redshift of around unity.
This means that $\xi(z=1)$ can be inferred quite
robustly, without much dependence on the rate of
evolution. Since the strength of clustering for
optical galaxies at $z=1$ is known to correspond to the
much smaller number of $r_0\simeq 2\mpcoh$
(e.g. Le F\`evre et al. 1996), we see that
radio galaxies at this redshift have a relative bias
parameter of close to 3.
This tendency for the relative bias to increase with
redshift probably arises partly because the high-redshift
sample members will be more powerful radio galaxies,
but also is likely to be related to the rareness of
such massive host galaxies at early times.

\sec{Formation of high-redshift galaxies}

The challenge now is to ask how these results can be
understood in current models for cosmological structure formation.
It is widely believed that the sequence of cosmological
structure formation was hierarchical, originating in 
a density power spectrum with increasing fluctuations on small scales.
The large-wavelength portion of this spectrum
is accessible to observation today through studies of
galaxy clustering in the linear and quasilinear regimes.
However, nonlinear evolution has effectively erased
any information on the initial spectrum for wavelengths
below about 1 Mpc. The most sensitive way of measuring
the spectrum on smaller scales is via the abundances of
high-redshift objects; the amplitude of fluctuations
on scales of individual galaxies governs the redshift
at which these objects first undergo gravitational collapse.
The small-scale amplitude also influences clustering, since rare
early-forming objects are strongly correlated,
as first realized by Kaiser (1984).

It will be especially interesting to apply these arguments about the
small-scale spectrum to a class of very early-forming galaxies
discussed at this meeting by Dunlop.
These are the red
optical identifications of 1-mJy radio galaxies, for which
deep absorption-line spectroscopy has proved 
that the red colours result
from a well-evolved stellar population, with a minimum
stellar age of 3.5 Gyr for 53W091 at
$z=1.55$ (Dunlop et al. 1996; Spinrad et al. 1997), and 4.0 Gyr for
53W069 at $z=1.43$ (Dey et al. 1998). Such ages push the
formation era for these galaxies back to extremely high
redshifts, and it is of interest to ask what level of small-scale
power is needed in order to allow this early formation.

\ssec{Press-Schechter apparatus}

The standard framework for interpreting the abundances
of high-redshift objects in terms of structure-formation models,
was outlined by Efstathiou \& Rees (1988). 
The formalism of Press \& Schechter (1974) gives a way of calculating
the fraction $F_c$ of the mass in the universe which has collapsed into objects
more massive than some limit $M$:
\japeqn
 F_c(>M,z) = 
 1 - {\rm erf}
\,\left[ {\delta_c \over \sqrt{2}\, \sigma(M)}\right].
\japeqn  
Here, $\sigma(M)$ is the rms fractional density contrast
obtained by filtering the linear-theory density field on the 
required scale. In practice, this filtering is usually performed
with a spherical `top hat' filter 
of radius $R$,  with a corresponding mass of $4 \pi \rho_b R^3/3 $,
where $\rho_b$ is the background density. 
The number $\delta_c$
is the linear-theory critical overdensity, which for a `top-hat'
overdensity undergoing spherical collapse is $1.686$ -- virtually
independent of $\Omega$. This form
describes numerical simulations very well (see e.g. 
Ma \& Bertschinger 1994). 
The main assumption is that the density field obeys Gaussian
statistics, which is true in most inflationary models.
Given some estimate of $F_c$, the number $\sigma(R)$
can then be inferred. Note that for rare objects this is a
pleasingly robust process: a large error in $F_c$ will give
only a small error in $\sigma(R)$, because the abundance is
exponentially sensitive to $\sigma$.

Total masses are of course ill-defined, and a better
quantity to use is the velocity dispersion.
Virial equilibrium for a halo of mass $M$ and proper radius $r$ demands
a circular orbital velocity of
\japeqn
V_c^2 = {GM \over r}
\japeqn 
For a spherically collapsed object this velocity  can be converted directly
into a Lagrangian comoving radius 
which contains the mass of the object within the virialization radius
(e.g. White, Efstathiou \& Frenk 1993):
\japeqn
R / \mpcoh= {2^{1/2}[V_c/100 \kms] \over  \Omega_m^{1/2}(1+z_c)^{1/2} f_c^{1/6}}.
\japeqn
Here,  $z_c$ is the redshift of virialization; $\Omega_m$ is
the {\it present\/} value of the matter density parameter;
$f_c$ is the density contrast at virialization
of the newly-collapsed object relative
to the background, which is adequately approximated by
\japeqn
f_c=178/\Omega_m^{0.6}(z_c),
\japeqn
with only a slight sensitivity to whether $\Lambda$ is non-zero
(Eke, Cole \& Frenk 1996).

For isothermal-sphere haloes, the velocity dispersion is
\japeqn
\sigma_v=V_c/\sqrt{2}.
\japeqn
Given a formation redshift of interest, and a velocity dispersion, there
is then a direct route to the Lagrangian radius from which the
proto-object collapsed.

\ssec{Abundances and masses of high-redshift objects}

In addition to the red mJy galaxies, two classes of high-redshift
object have been used recently to set constraints on
the small-scale power spectrum at high redshift:

\noindent{\bf (1) Damped Lyman-$\alpha$ systems}
\quad
Damped Lyman-$\alpha$ absorbers are systems with HI column densities greater than
$\sim 2\times 10^{24}\; \rm m^{-2}$ (Lanzetta et al. 1991).
If the fraction of baryons in
the virialized dark matter halos equals the global value $\Omega_{\ss B}$,
then data on these systems can be used to infer the total fraction of matter that has 
collapsed into bound structures at high redshifts (Ma \& Bertschinger 1994,
Mo \& Miralda-Escud\'{e} 1994; Kauffmann \& Charlot 1994;
Klypin et al. 1995). The highest
measurement at $\langle z \rangle \simeq 3.2$ implies 
$\Omega_{\ss HI}\simeq 0.0025h^{-1}$
(Lanzetta et al. 1991; Storrie-Lombardi, McMahon \& Irwin 1996).
If $\Omega_{\ss B}h^2 =0.02$ is adopted, as a compromise between the lower
Walker et al. (1991) nucleosynthesis estimate and the more recent
estimate of 0.025 from Tytler et al. (1996), then
\japeqn
F_c = {\Omega_{\ss HI}\over \Omega_{\ss B}} \simeq 0.12h
\japeqn
for these systems.
In this case alone, an explicit value of $h$ is required in order to
obtain the collapsed fraction; $h=0.65$ is assumed.

The photoionizing background prevents virialized gaseous
systems with circular
velocities of less than about $50 \kms$ from cooling efficiently,
so that they cannot contract to the high density contrasts 
characteristic of galaxies (e.g. Efstathiou 1992). 
Mo \& Miralda-Escud\'{e} (1994) used the circular velocity range
50  -- $100\kms$ ($\sigma_v=35$ -- $70\kms$) to model the damped
Lyman alpha systems. Reinforcing the photoionization argument,
detailed hydrodynamic simulations imply that the absorbers
are not expected to be associated with very massive dark-matter haloes 
(Haehnelt, Steinmetz \& Rauch 1997). This assumption is consistent with the
rather low luminosity galaxies detected in association with the absorbers
in a number of cases (Le Brun et al. 1996).

\noindent{\bf (2) Lyman-limit galaxies}
\quad
Steidel et al. (1996) identified star-forming galaxies between
$z=3$ and 3.5 by looking for objects with a spectral
break redwards of the $U$ band. 
The treatment of these Lyman-limit galaxies in this paper is 
similar to that of Mo \& Fukugita (1996), who compared
the abundances of these objects to predictions from
various models.
Steidel et al. give the comoving density of their galaxies as
\japeqn
N(\Omega=1) \simeq 10^{-2.54} \; (\mpcoh)^{-3}.
\japeqn
This is a high number density, comparable to that of
$L^*$ galaxies in the present Universe. The mass of
$L^*$ galaxies corresponds to collapse of a Lagrangian
region of volume $\sim 1\,\rm Mpc^3$, so the collapsed
fraction would be a few tenths of a per cent if the
Lyman-limit galaxies had similar masses.

Direct dynamical determinations of these masses
are still lacking in most cases. Steidel et al. attempt to
infer a velocity width by looking at the equivalent
width of the C and Si absorption lines. These are
saturated lines, and so the equivalent width is
sensitive to the velocity dispersion; values in the
range 
\japeqn
\sigma_v\simeq 180 - 320 \kms
\japeqn 
are implied. These numbers may measure velocities
which are not due to bound material, in which case
they would give an upper limit to $V_c/\sqrt{2}$ for the
dark halo. A more recent measurement of
the velocity width of the H$\alpha$ emission line in one of these objects
gives a dispersion of closer to $100 \kms$ (Pettini, private
communication), consistent with the median velocity
width for Ly$\alpha$ of $140\kms$ measured in similar
galaxies in the HDF (Lowenthal et al. 1997).
Of course, these figures could underestimate the total velocity
dispersion, since they are dominated by emission from the central regions only.
For the present, the range of values $\sigma_v = 100$ to $320 \kms$
will be adopted, and
the sensitivity to the assumed velocity will be indicated.
In practice, this uncertainty in the velocity does
not produce an important uncertainty in the conclusions.

\noindent{\bf (3) Red radio galaxies}
\quad
Two extremely red galaxies were found at $z=1.43$ and 1.55, 
over an area $1.68\times 10^{-3}\; \rm sr$, so a minimal
comoving density is from one galaxy in this redshift range:
\japeqn
N(\Omega=1) \gs 10^{-5.87} \; (\mpcoh)^{-3}.
\japeqn
This figure is comparable to
the density of the richest Abell clusters, and is thus in reasonable
agreement with the discovery that rich high-redshift
clusters appear to contain radio-quiet examples
of similarly red galaxies (Dickinson 1995).

Since the velocity dispersions of these galaxies are
not observed, they must be inferred indirectly. This
is possible because of the known present-day Faber-Jackson
relation for ellipticals. For 53W091, the large-aperture
absolute magnitude is
\japeqn
M_V(z=1.55\mid \Omega=1) \simeq -21.62 -5 \log_{10} h
\japeqn
(measured direct in the rest frame).
According to Solar-metallicity  spectral synthesis models, 
this would be expected to fade
by about 0.9 mag. between $z=1.55$ and the present, for
an $\Omega=1$ model of present age 14 Gyr
(note that Bender et al. 1996 have observed a shift
in the zero-point of the $M-\sigma_v$ relation out to $z=0.37$
of a consistent size).
If we compare these numbers with the $\sigma_v$ -- $M_V$
relation for Coma ($m-M=34.3$ for $h=1$) taken from
Dressler (1984), this predicts velocity dispersions in the range
\japeqn
\sigma_v= 222 \; {\rm to}\; 292 \; \kms.
\japeqn
This is a very reasonable range for a giant elliptical,
and it adopted in the following analysis.

Having established an abundance and an equivalent circular velocity
for these galaxies, the treatment of them will differ in one
critical way from the Lyman-$\alpha$ and Lyman-limit galaxies.
For these, the normal Press-Schechter approach assumes 
the systems under study to be newly born. For
the Lyman-$\alpha$ and Lyman-limit galaxies, 
this may not be a bad approximation,
since they are evolving rapidly and/or display high levels of star-formation
activity. For the radio galaxies, conversely, 
their inactivity suggests that they may have existed as
discrete systems at redshifts much higher than $z\simeq 1.5$.
The strategy will therefore be to
apply the Press-Schechter machinery at some unknown formation
redshift, and see what range of redshift gives a consistent
degree of inhomogeneity.

\japfig{3}{3}
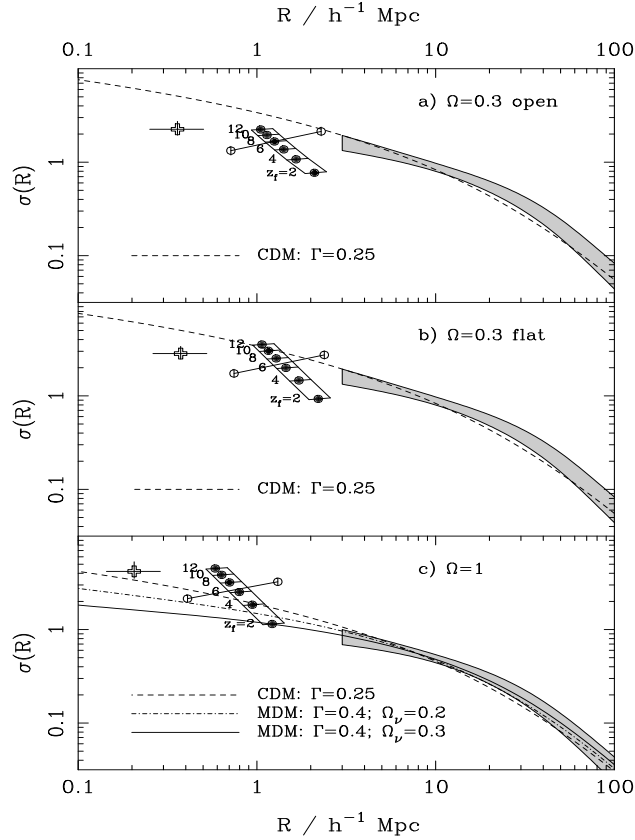
{The present-day linear 
fractional rms fluctuation in density averaged in
spheres of radius $R$. The data points are
Lyman-$\alpha$ galaxies (open cross) and
Lyman-limit galaxies (open circles)
The diagonal band with solid points shows red radio
galaxies with assumed collapse redshifts 2, 4, \dots 12.
The vertical error bars show the effect of a change in abundance by a factor 2.
The horizontal errors correspond to different choices for the
circular velocities of the dark-matter haloes that host the galaxies.
The shaded region at large $R$ gives the results inferred
from galaxy clustering.
The lines show CDM and MDM predictions,
with a large-scale normalization of $\sigma_8=0.55$ for $\Omega=1$
or $\sigma_8=1$ for the low-density models.}

\sec{The small-scale fluctuation spectrum}

\ssec{The empirical spectrum}

Fig. 3 shows the $\sigma(R)$ data which result
from the Press-Schechter analysis, for three
cosmologies. The $\sigma(R)$ numbers measured at various
high redshifts have been translated to $z=0$ using the
appropriate linear growth law for density perturbations.

The open symbols give the results for the
Lyman-limit (largest $R$) and Lyman-$\alpha$ (smallest $R$)
systems. The approximately horizontal error bars show the effect of the
quoted range of velocity dispersions for a fixed
abundance; the vertical
errors show the effect of changing the abundance by a factor 2 at fixed
velocity dispersion.
The locus implied by the red radio galaxies sits
in between. The different points show the effects of
varying collapse redshift: $z_c=2, 4, \dots, 12$
[lowest redshift gives lowest $\sigma(R)$]. Clearly, collapse
redshifts of 6 -- 8 are favoured for consistency
with the other data on high-redshift galaxies, independent
of theoretical preconceptions and independent of the
age of these galaxies. This level of power
($\sigma[R]\simeq 2$ for $R\simeq 1 \mpcoh$) is also
in very close agreement with the level of power required to
produce the observed structure in the Lyman alpha forest
(Croft et al. 1997), so there is a good case to be made that
the fluctuation spectrum has now been measured in a 
consistent fashion down to below $R\simeq 1\mpcoh$.

The shaded region at larger $R$ shows the results
deduced from clustering data (Peacock 1997). 
It is clear an $\Omega=1$ universe requires the power spectrum
at small scales to be higher than would be expected on the basis of an
extrapolation from the large-scale spectrum. Depending on assumptions
about the scale-dependence of bias, such a `feature'
in the linear spectrum  may also
be required in order to satisfy the small-scale present-day 
nonlinear galaxy clustering (Peacock 1997).
Conversely, for low-density models, the empirical small-scale
spectrum appears to match reasonably smoothly onto the large-scale data.

Fig. 3 also compares the empirical data with various physical power
spectra. A CDM model (using the transfer 
function of Bardeen et al. 1986) with shape parameter
$\Gamma=\Omega h=0.25$ is shown as a reference for all models.
This has approximately the correct level of small-scale
power, but significantly  over-predicts intermediate-scale clustering, as
discussed in Peacock (1997). The empirical LSS shape is better
described by MDM with $\Omega h\simeq 0.4$ and $\Omega_\nu\simeq 0.3$.
This is the lowest curve in Fig. 1c, reproduced from the
fitting formula of Pogosyan \& Starobinsky (1995; see also
Ma 1996). However, this curve fails to
supply the required small-scale power,
by about a factor 3 in $\sigma$; lowering
$\Omega_\nu$ to 0.2 still leaves a very large discrepancy.
This conclusion is in agreement with e.g. Mo \& Miralda-Escud\'e (1994),
Ma \& Bertschinger (1994), but conflicts slightly with
Klypin et al. (1995), who claimed that the $\Omega_\nu=0.2$
model was acceptable.
This difference arises partly because Klypin et al. adopt a
lower value for $\delta_c$ (1.33 as against 1.686 here), and
also because they adopt the high normalization of
$\sigma_8=0.7$; the net effect of these changes is to boost
the model relative to the small-scale data by a factor of 1.6,
which would allow marginal consistency for the $\Omega_\nu=0.2$
model. MDM models do allow a higher normalization than
the conventional figure of $\sigma_8=0.55$, partly because of
the very flat small-scale spectrum, and also because of the
effects of random neutrino velocities. However, such shifts
are at the 10 per cent level (Borgani et al. 1997a, 1997b), 
and $\sigma_8=0.7$ would probably
still give a cluster abundance in excess of observation.
The consensus of more recent modelling is that even 
$\Omega_\nu=0.2$ MDM is deficient in small-scale power
(Ma et al. 1997; Gardner et al. 1997).

All the models in Fig. 1 assume $n=1$; in fact, consistency with the COBE
results for this choice of $\sigma_8$ 
and $\Omega h$ requires a significant
tilt for flat CDM models, $n\simeq 0.9$ (whereas open CDM
models require $n$ substantially above unity).
Over the range of scales probed by LSS,
changes in $n$ are largely degenerate with changes in $\Omega h$,
but the small-scale power is more sensitive to tilt
than to $\Omega h$. Tilting the $\Omega=1$ models is
not attractive, since it increases the tendency for model
predictions to lie below the data. However,
a tilted low-$\Omega$ flat CDM model would agree moderately
well with the data on all scales, with the exception of the
`bump' around $R\simeq 30 \mpcoh$. Testing the reality of this
feature will therefore be an important task for future
generations of redshift survey.

\ssec{Collapse redshifts and ages}

Are the collapse redshifts 
inferred above consistent with the
age data on the red radio galaxies? First bear in mind that
in a hierarchy some of the stars in a galaxy will inevitably
form in sub-units before the epoch of collapse.
At the time of final collapse,
the typical stellar age will be some fraction $\alpha$ of
the age of the universe at that time:
\japeqn
{\rm age} = t(z_{\rm obs}) - t(z_c) + \alpha t(z_c).
\japeqn
We can rule out $\alpha=1$ (i.e. all stars forming in small subunits  just
after the big bang). For present-day
ellipticals, the tight colour-magnitude relation
only allows an approximate doubling of the mass
through mergers since the termination of star formation
(Bower at al. 1992). This corresponds to $\alpha\simeq 0.3$
(Peacock 1991). A non-zero $\alpha$ just corresponds to
scaling the collapse redshift as
\japeqn
{\rm apparent}\ (1+z_c)\propto (1-\alpha)^{-2/3},
\japeqn
since $t\propto (1+z)^{-3/2}$
at high redshifts for all cosmologies.
For example, a galaxy which collapsed at $z=6$ would have
an apparent age corresponding to a collapse redshift of 7.9 for $\alpha=0.3$.

Converting the ages for the galaxies to an apparent collapse
redshift depends on the cosmological model, but particularly on $H_0$.
Some of this uncertainty may be circumvented by fixing the age of the
universe. After all, it is of no interest to ask about formation
redshifts in a model with e.g. $\Omega=1$, $h=0.7$ when the whole
universe then has an age of only 9.5 Gyr. If $\Omega=1$ is to be tenable
then either $h<0.5$ against all the evidence or there must be an error
in the stellar evolution timescale. If the stellar timescales
are wrong by a fixed factor, then these two possibilities
are degenerate. It therefore makes sense to measure galaxy ages
only in units of the age of the universe -- or, equivalently, 
to choose freely an apparent Hubble constant which gives the
universe an age comparable to that inferred for globular clusters.
In this spirit, Fig. 4 gives
apparent ages as a function of effective collapse redshift for
models in which the age of the universe is forced to be 14 Gyr
(e.g. Jimenez et al. 1996).

\japfig{5}{1}
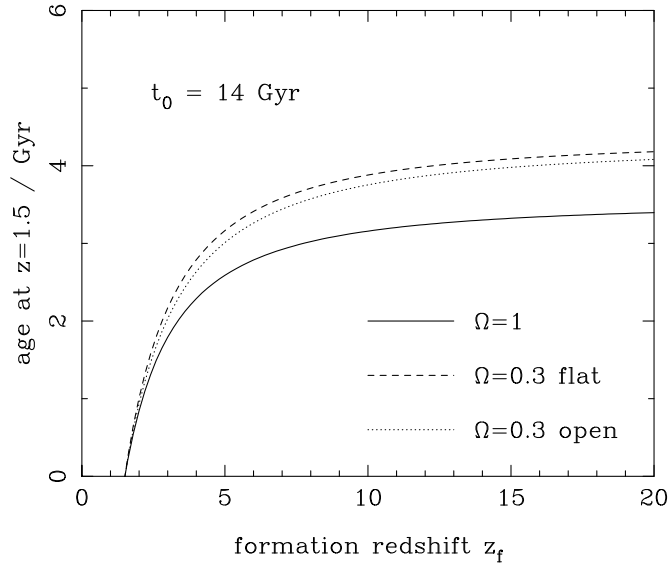
{The age of a galaxy at $z=1.5$, as a function of its
collapse redshift (assuming an instantaneous burst of star formation).
The various lines show $\Omega=1$ [solid]; open $\Omega=0.3$ [dotted];
flat $\Omega=0.3$ [dashed]. In all cases, the present
age of the universe is forced to be 14 Gyr.}

This plot shows that the ages of the red radio galaxies are
not permitted very much freedom. 
Formation redshifts in the range 6 to 8
predict an age of close to 3.0 Gyr for $\Omega=1$,
or 3.7 Gyr for low-density models, irrespective of
whether $\Lambda$ is nonzero.
The age-$z_c$ relation is rather flat, and this gives
a robust estimate of age once we have some idea of $z_c$
through the abundance arguments.
Conversely, it is almost impossible to determine the
collapse redshift reliably from the spectral
data, since a very high precision would be required both
in the age of the galaxy and in the age of the universe.

What conclusions can then be reached about allowed cosmological models?
If we take an apparent $z_c=8$ from the power-spectrum arguments, then
the apparent minimum age of $>4$ Gyr for 53W069 can very nearly be
satisfied in both low-density models
(a current age of 14.5 Gyr would be required), but is unattainable for $\Omega=1$.
In the high-density case, a current age of 17.6 Gyr would be required
to attain the required age for $z_c=8$;
this requires a Hubble constant of $h=0.38$.
As argued above, this conclusion is highly insensitive
to the assumed value of $z_c$.
If the true value of $h$ does turn out to be close to 0.5,
then it might be argued that $\Omega=1$ is consistent with the
data, given realistic uncertainties. The ages for the low-density
models would in this case be large by comparison with the
observed radio-galaxy ages. However, 
the ages obtained by modelling spectra with a single burst
can only be lower limits to the true age for the bulk of the stars;
we could easily be observing an even older burst which is made bluer by
a little recent star formation.
A low $h$ measurement would therefore not rule
out low-density models.

\japfig{4}{1}
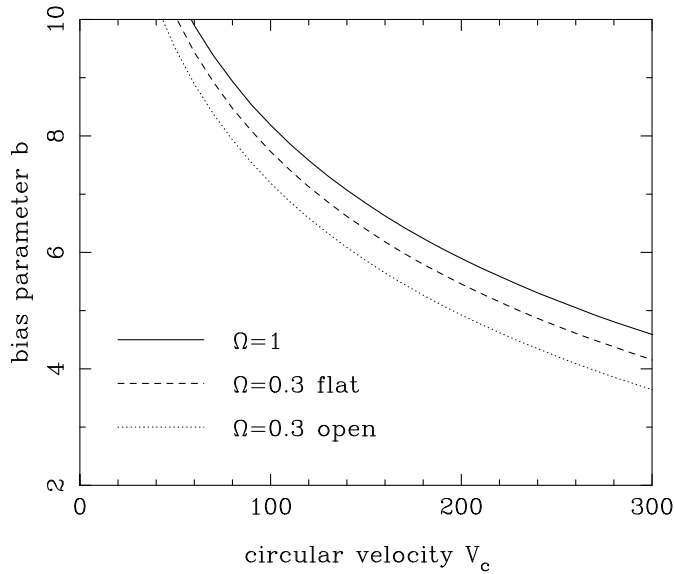
{The bias parameter at $z=3.2$ predicted for the
Lyman-limit galaxies, as a function of their assumed 
circular velocity. Dotted line shows $\Omega=0.3$ open;
dashed line is $\Omega=0.3$ flat; solid line is $\Omega=1$.
A substantial bias in the region of $b\simeq 6$ is predicted
rather robustly.}

\sec{Biased clustering at high redshifts}

\ssec{Predictions from the power spectrum}

An interesting aspect of these results is that the
level of power on 1-Mpc scales is only moderate:
$\sigma(1\mpcoh)\simeq 2$. At $z\simeq 3$, the
corresponding figure would have been much lower,
making systems like the Lyman-limit galaxies rather
rare. For Gaussian fluctuations, as assumed in the
Press-Schechter analysis, such systems will be
expected to display spatial correlations which are
strongly biased with respect to the underlying mass.
The linear bias parameter depends on the rareness of
the fluctuation and the rms of the underlying field as
\japeqn
b=1+{\nu^2-1\over \nu\sigma}= 1+ {\nu^2-1\over \delta_c}
\japeqn
(Kaiser 1984; Cole \& Kaiser 1989; Mo \& White 1996),
where $\nu = \delta_c/\sigma$, and $\sigma^2$ is the
fractional mass variance at the redshift of interest.

In this analysis, $\delta_c=1.686$ is assumed. 
Variations in this number of order 10 per cent have
been suggested by authors who have studied the
fit of the Press-Schechter model to numerical data.
These changes would merely scale $b-1$ by a small amount;
the key parameter is $\nu$, which is set entirely by
the collapsed fraction. For the Lyman-limit galaxies,
typical values of this parameter are $\nu\simeq 3$,
and it is clear that very substantial values of bias
are expected, as illustrated in Figure 5.

This diagram shows how the predicted bias parameter
varies with the assumed circular velocity, for a number density of
galaxies fixed at the level observed by Steidel et al. (1996).
The sensitivity to cosmological parameter is only
moderate; at $V_c=200\kms$, the predicted bias is $b\simeq 4.6$, 5.5, 5.8
for the open, flat and critical models respectively.
These numbers scale approximately as $V_c^{-0.4}$, and
$b$ is within 20 per cent of 6 for most plausible
parameter combinations.
Strictly, the bias values determined here are upper
limits, since the numbers of collapsed haloes of this
circular velocity could in principle greatly exceed the
numbers of observed Lyman-limit galaxies. However, the
undercounting would have to be substantial: increasing the
collapsed fraction by a factor 10 reduces the implied bias
by a factor of about 1.5. A substantial bias seems
difficult to avoid, as has been pointed out in the context of CDM
models by Baugh, Cole \& Frenk (1997).

\ssec{Clustering of Lyman-limit galaxies}

These calculations are relevant to the
recent detection by Steidel et al. (1997) of strong
clustering in the population of Lyman-limit galaxies at $z\simeq 3$.
The evidence takes the form of a redshift histogram binned
at $\Delta z=0.04$ resolution over a field $8.7' \times 17.6'$ in extent.
For $\Omega=1$ and $z=3$, this probes the density field using a cell with dimensions
\japeqn
{\rm cell} = 15.4 \times 7.6 \times 15.0 \; [\mpcoh]^3.
\japeqn
Conveniently, this has a volume equivalent to a sphere of radius
$7.5 \mpcoh$, so it is easy to measure the bias directly by reference
to the known value of $\sigma_8$. Since the degree of bias is large,
redshift-space distortions from coherent infall are small;
the cell is also large enough that the distortions of small-scale
random velocities at the few hundred $\kms$ level are also small.
Using the model of equation (11) of Peacock (1997) for the
anisotropic redshift-space power spectrum and integrating over
the exact anisotropic window function, the above simple
volume argument is found to be accurate  to a few per cent for reasonable
power spectra:
\japeqn
\sigma_{\rm cell} \simeq b(z=3) \; \sigma_{7.5}(z=3),
\japeqn
defining the bias factor at this scale. The results of
Mo \& White (1996) suggest that the scale-dependence of bias should be weak.

In order to estimate $\sigma_{\rm cell}$, simulations of 
synthetic redshift histograms were made,
using the method of Poisson-sampled
lognormal realizations described by Broadhurst, Taylor \& Peacock (1995):
using a $\chi^2$ statistic to quantify the nonuniformity of the
redshift histogram, it appears that $\sigma_{\rm cell}\simeq 0.9$
is required in order for the field of Steidel et al. (1997) to be typical.
It is then straightforward to obtain the bias parameter since, for a 
present-day correlation function $\xi(r)\propto r^{-1.8}$,
\japeqn
\sigma_{7.5}(z=3)=\sigma_8 \times [8/7.5]^{1.8/2} \times 1/4 \simeq 0.146,
\japeqn
implying
\japeqn
b(z=3\mid\Omega=1)\simeq 0.9/0.146 \simeq 6.2.
\japeqn
Steidel et al. (1997) use a rather different analysis which concentrates
on the highest peak alone, and obtain a minimum bias of 6, with a preferred
value of 8. They use the Eke et al. (1996) value of $\sigma_8=0.52$, which
is on the low side of the published range of estimates. Using $\sigma_8=0.55$
would lower their preferred $b$ to 7.6.
Note that, with both these methods, it is much easier to rule out
a low value of $b$ than a high one; given a single field, it is
possible that a relatively `quiet' region of space has been sampled,
and that much larger spikes remain to be found elsewhere.

Having arrived at a figure for bias if $\Omega=1$, it is easy to
translate to other models, since $\sigma_{\rm cell}$ is observed,
independent of cosmology. For low $\Omega$ models, the cell volume
will increase by a factor $[S_k^2(r)\, dr]/[S_k^2(r_1)\, dr_1]$; comparing with
present-day fluctuations on this larger scale will tend to increase
the bias. However, for low $\Omega$, two other effects increase the
predicted density fluctuation at $z=3$: the cluster constraint
increases the present-day fluctuation by a factor $\Omega^{-0.56}$, and
the growth between redshift 3 and the present will be less than
a factor of 4. Applying these corrections gives
\japeqn
{ b(z=3 \mid \Omega=0.3) \over b(z=3 \mid \Omega=1) } =
\left\{ {0.42\ ({\rm open}) \atop 0.60\ \rlap{({\rm flat})}
\phantom{({\rm open})}} \right. ,
\japeqn
which suggests an approximate scaling as $b\propto \Omega^{0.72}$ (open)
or $\Omega^{0.42}$ (flat). The significance of this observation is thus
to provide the first convincing proof for the reality of galaxy bias: for
$\Omega\simeq 0.3$, bias is not required in the present universe,
but we now see that $b>1$ is needed at $z=3$ for all reasonable
values of $\Omega$.

Comparing these bias values with Fig. 5, we see that
the observed value of $b$ is quite close to the prediction
in the case of $\Omega=1$
-- suggesting that the simplest interpretation of these
systems as collapsed rare peaks may well be roughly correct.
Indeed, for high circular velocities
there is a danger of exceeding the predictions, and it would
create something of a difficulty for high-density models if
a velocity as high as $V_c\simeq 300 \kms$ were to be established as
typical of the Lyman-limit galaxies.
For low $\Omega$, the `observed' bias falls faster than the
predictions, so there is less danger of conflict. For a circular
velocity of $200 \kms$, we would need to say that the collapsed fraction
was underestimated by roughly a factor 10
(i.e. increase the values of $\sigma$ in Fig. 1 by a factor
of about 1.5) in order to lower the
predicted bias sufficiently, either by postulating that the
conversion from velocity to $R$ is systematically in error, or
by suggesting that there may be many haloes which are not detected
by the Lyman-limit search technique. It is hard to argue that
either of these possibilities are completely ruled out.
Nevertheless, we have reached the paradoxical conclusion that
the observed large-amplitude clustering at $z=3$ is more naturally
understood in an $\Omega=1$ model, whereas one might have expected the
opposite conclusion.

\ssec{Clustering of high-redshift AGN}

The strength of clustering for Lyman-limit galaxies
fits in reasonably well with what is known about
clustering of AGN. The earlier sections have argued
for $r_0\simeq 6.5\mpcoh$ for radio galaxies at
$z\simeq 1$. An almost identical correlation length
has been measured for radio-quiet QSOs at $\langle z \rangle
\simeq 1.5$ (Shanks \& Boyle 1994; Croom \& Shanks 1996).
These values are much larger than the clustering of
optically-selected galaxies at $z\simeq 1$, but this
is not unreasonable, since imaging of QSO hosts
reveals them to be several-$L^*$ objects, comparable
in stellar mass to radio galaxies
(e.g. Dunlop et al. 1993; Taylor et al. 1996).
It is plausible that the clustering of these massive
galaxies at $z\simeq 1$ will be enhanced through
exactly the same mechanisms that enhances the clustering
of Lyman-limit galaxies at $z\simeq 3$. Of course, this
does not rule out more complex pictures based on ideas
such as close interactions in rich environments being
necessary to trigger AGN. However, as emphasised above,
the mass and rareness of these objects sets a
{\it minimum\/} level of bias. It is to be expected
that this bias will increase at higher redshifts,
and so one would not expect quasar clustering to
decline at higher redshifts. Indeed, it has
been claimed that $\xi$ either stays constant at the highest
redshifts (Andreani \& Cristiani 1992; Croom \& Shanks 1996),
or even increases (Stephens et al. 1997).

\sec{The global picture of galaxy formation}

This paper has advanced the view that there is
a good degree of consistency between the emerging
data on both the abundances and the clustering of
a variety of high-redshift galaxies.
It is especially interesting to note that it has been
possible to construct a consistent picture which
incorporates both the large numbers of star-forming
galaxies at $z\ls 3$ and the existence of old systems
which must have formed at very much larger redshifts.
A recent conclusion from the numbers of Lyman-limit
galaxies and the star-formation rates seen at $z\simeq 1$
has been that the global history of star formation
peaked at $z\simeq 2$ (Madau et al. 1996). This leaves open two possibilities
for the very old systems: either they are the rare precursors
of this process, and form unusually early, or they are
a relic of a second peak in activity at higher redshift,
such as is commonly invoked for the origin of all
spheroidal components.
While such a bimodal history of star formation
cannot be rejected, the rareness of the
red radio galaxies indicates that there is no difficulty
with the former picture. This can be demonstrated quantitatively
by integrating the total amount of star formation at high redshift.
According to Madau et al., The star-formation rate at $z=4$ is
\japeqn
\dot \rho_* \simeq 10^{7.3}h\; M_\odot\,{\rm Gyr}^{-1}\, {\rm Mpc}^{-3},
\japeqn
declining roughly as $(1+z)^{-4}$. This is probably a underestimate
by a factor of at least 3, as indicated by suggestions of dust in
the Lyman-limit galaxies (Pettini et al. 1997), and by the prediction
of Pei \& Fall (1995), based on high-$z$ element abundances.
If we scale by a factor 3, and integrate to find the total density
in stars produced at $z>6$, this yields
\japeqn
\rho_*(z_{\rm f}>6) \simeq 10^{6.2} M_\odot\,{\rm Mpc}^{-3}.
\japeqn
Since the mJy galaxies have a density of $10^{-5.87}h^3 {\rm Mpc}^{-3}$
and stellar masses of order $10^{11}\, M_\odot$, there is clearly no
conflict with the idea that these galaxies are 
the first stellar systems of $L^*$ size which form
en route to the general era of star and galaxy formation.

The data on the abundances and clustering of both
radio-loud and radio-quiet galaxies at high redshift
thus appear to be in good quantitative agreement with
the expectation of models in which structure formation
proceeds through hierarchical merging of haloes of dark matter.
Furthermore, the existing
data yield an empirical measurement of the fluctuation spectrum
which is required on sub-Mpc scales.
In general, this small-scale spectrum is close to
what would be expected from an extrapolation of
LSS measurements, but there are deviations in detail:
$\Omega=1$ places the small-scale data somewhat above
the LSS extrapolation, whereas open low-$\Omega$ models suffer from the
opposite problem; low-$\Omega$ $\Lambda$-dominated models
fare somewhat better. These last models also do
reasonably well if the dark matter is assumed to be pure
CDM, normalized to COBE (whereas open models do not). 
The main difficulties for $\Lambda$CDM lie in the
shape of the large-scale power spectrum measured from the APM survey,
and in geometrical diagnostics such as the supernova
Hubble diagram. The fact that the  $\Lambda$CDM model
provides the best match with the empirical small-scale
spectrum should encourage further critical examination
of these objections.
The subject of structure formation stands at a critical point:
either we are close to having a `standard model' for
galaxy formation and clustering, or we may
have to accept that radical new ideas are needed.
At the current rate of observational progress, the
verdict should not be very far away.

\section*{Acknowledgements}
This paper draws on unpublished collaborative work
with Alison Stirling (sections 1 -- 2) and
James Dunlop, Raul Jimenez, Ian Waddington, Hy Spinrad,
Daniel Stern, Arjun Dey \& Rogier Windhorst (sections 3 -- 6).

\end{document}